\documentclass[sigconf]{acmart}
\setcopyright{acmcopyright}
\copyrightyear{2024}
\acmYear{2024}
\acmDOI{TBD}
\acmConference[]{}{}

\usepackage{amsmath,amsfonts}
\usepackage{textcomp}
\usepackage{xcolor}
\usepackage{hyperref}
\usepackage{graphicx, color}
\begin{document}

\title{Model-Checking in the Loop Model-Based Testing for Automotive Operating Systems}
\author{Toshiaki Aoki}
\email{toshiaki@jaist.ac.jp}
\affiliation{%
  \institution{JAIST}
  \streetaddress{1-1 Asahidai}
  \city{Nomi}
  \state{Ishikawa}
  \country{JAPAN}
  \postcode{923-1292}
}
\author{Aritoshi Hata}
\email{aritoshi.hata.j2d@jp.denso.com}
\author{Kazusato Kanamori}
\email{kazusato.kanamori.j8k@jp.denso.com}
\author{Satoshi Tanaka}
\email{satoshi.tanaka.j7d@jp.denso.com}
\affiliation{%
  \institution{DENSO CORPORATION}
  \streetaddress{1-1, Showa-cho}
  \city{Kariya}
  \state{Aichi}
  \country{JAPAN}
  \postcode{448-8661}
}
\author{Yuta Kawamoto}
\email{yuta.kawamoto.j8f@jpgr.denso.com}
\author{Yasuhiro Tanase}
\email{yasuhiro.tanase.j7g@jpgr.denso.com}
\author{Masumi Imai}
\email{masumi.imai.j3w@jpgr.denso.com}
\affiliation{%
  \institution{DENSO CREATE INC.}
  \streetaddress{NAGOYA FUSHIMI K-SQUARE Bldg., 2-14-19, Nishiki Naka-ku}
  \city{Nagoya}
  \state{Aichi}
  \country{JAPAN}
  \postcode{460-0003}
}
\author{Fumiya Shigemitsu}
\email{f-shigemitsu@esol.co.jp}
\author{Masaki Gondo}
\email{gondou@esol.co.jp}
\affiliation{%
  \institution{eSOL Co.,Ltd}
  \streetaddress{Harmony Tower, 1-32-2, Honcho}
  \city{Nakano-ku}
  \state{Tokyo}
  \country{JAPAN}
  \postcode{164-8721}
}
\author{Tomoji Kishi}
\email{kishi@waseda.jp}
\affiliation{%
  \institution{Waseda University}
  \streetaddress{3-4-1 Ookubo}
  \city{Shinjuku-ku}
  \state{Tokyo}
  \country{JAPAN}
  \postcode{169-8555}
}
\renewcommand{\shortauthors}{T. Aoki et al.}

\begin{abstract}
While vehicles have primarily been controlled through mechanical means in
years past, an increasing number of embedded control systems are being
installed and used, keeping pace with advances in electronic control
technology and performance. Automotive systems consist of multiple
components developed by a range of vendors. To accelerate developments in
embedded control systems, industrial standards such as AUTOSAR are being
defined for automotive systems, including the design of operating system
and middleware technologies. Crucial to ensuring the safety of automotive
systems, the operating system is foundational software on which many
automotive applications are executed. In this paper, we propose an
integrated model-based method for verifying automotive operating systems;
our method is called Model-Checking in the Loop Model-Based Testing
(MCIL-MBT). In MCIL-MBT, we create a model that formalizes specifications
of automotive operating systems and verifies the specifications via
model-checking. Next, we conduct model-based testing with the verified
model to ensure that a specific operating system implementation conforms
to the model. These verification and testing stages are iterated over
until no flaws are detected. Our method has already been introduced to an
automotive system supplier and an operating system vendor. Through our
approach, we successfully identified flaws that were not detected by
conventional review and testing methods.
\end{abstract}

\keywords{model-checking, model-based testing, automotive OS}
\maketitle
\section{Introduction}
The reliability and safety of automotive systems represent an increasingly
important concern in our society. Although vehicles have largely been
controlled through mechanical means in the past, many electronic embedded
control systems are now utilized to keep pace with advances in electronic
control technology and performance. While the use of such electronic
control technology makes for more convenient and safer vehicles,
electronic parts unfortunately introduce problems of reliability and
safety due to increasingly complex and larger-scale interoperable software
components. In particular, highly electronic automotive systems have
received increased attention with respect to their reliability and
safety. For example, in 2010, electronic throttle systems were suspected
of having issues and therefore inspected by the National Highway Traffic
Safety Administration (NHTSA) and the National Aeronautics and Space
Administration (NASA) agencies in the United States based on unintended
acceleration problems in Toyota model cars \cite{UIAP}. According to such
concerns regarding reliability and safety, functional safety standards for
automotive systems were proposed by the International Organization for
Standardization (ISO) \cite{IEC61508,ISO26262}.

With safety and reliability in mind, we focus on the operating system
(OS), foundational software running on any computer system, including
present-day automotive systems. The quality of an OS is of critical
importance since it serves as such foundational software on top of which
other software executes. With various automotive systems built on top of
an OS, the quality of the OS directly affects those other software
systems. Further, such software systems are safety-critical systems in
which fatal accidents may occur if flaws in the software or underlying OS
exist. Given this, we have focused our work on automotive OS verification
to ensure the reliability and safety of automotive systems. In our
previous work, we verified a commercial OS developed by Renesas
Electronics \cite{JAIST5}, which conforms to {\it de facto} standard {\it
  OSEK/VDX} \cite{OSEK}. At present, it has been inherited by AUTOSAR,
with an automotive OS standard called {\it AUTOSAR OS} being developed
\cite{AUTOSAR}. The task management functions of AUTOSAR OS are the same
as those in the original OSEK/VDX, though new functionality has been
incorporated for protection and support for multi-core
processors. Conventional priority-based scheduling is used in AUTOSAR OS
because such an approach focuses on real-time systems; however, many-core
processors are increasingly being used for automotive systems, which,
given their scale and complexity, are causing conventional priority-based
scheduling to no longer be suitable.

As more flexible and scalable scheduling is required, a new automotive OS
standard called the {\it Adaptive AUTOSAR OS} \cite{AAUTOSAR} is being
developed. This Adaptive AUTOSAR OS is based on the Portable Operating
System Interface (POSIX) family of standards that employs flexible
scheduling techniques involving migration between processor cores. Given
these evolving circumstances, the scheduling mechanisms of automotive OS
implementations have become even more complicated, causing their
development and verification to be challenging. Note here that since the
basic scheduling scheme employed by the AUTOSAR OS, which was inherited
from OSEK/VDX, differs from that of the Adaptive AUTOSAR OS, the former is
referred to as the {\it Classic AUTOSAR OS}. In this paper, we will use
these two terms (i.e., Adaptive and Classic) to refer to these two
implementations; further, when we write just AUTOSAR OS, we are referring
to both the Adaptive and Classic AUTOSAR OS implementations. Also note
that a concurrent unit is called a {\it task} in the Classic AUTOSAR OS,
whereas it is called a {\it thread} in the Adaptive AUTOSAR OS.  For
simplicity, we use the term task throughout this paper to refer to either
a task or a thread.

Although the AUTOSAR OS standard documentation primarily describes the
roles and behavior of the OS application programming interfaces (APIs),
including return values, error codes, and so on, an exact mechanism for
determining how a task is to be dispatched is not described. Instead, only
execution examples of tasks are shown, and it is impossible to deduce the
exact scheduling behavior from such examples. These ambiguities make it
difficult to check whether a developed OS conforms to the standard and may
also introduce discrepancies between the OS development side and users. In
general, when we verify whether the scheduling of tasks conforms to the
standard, we test it against execution sequences of tasks; however, it is
difficult to cover all possible variations of execution sequences because
the number of variations can be huge.

Given the above, in this paper, we propose an integrated method called
{\it Model-Checking in the Loop Model-Based Testing} ({\it MCIL-MBT})
that combines both model-checking and specific model-based testing stages
to verify automotive OS implementations, in particular, their
scheduling. The fundamental intent of MCIL-MBT is to provide a means of
conducting acceptance testing. The target automotive OS is treated as a
black box in that the actual code that implements the OS cannot be
observed or analyzed; instead, we can feed inputs to the system and verify
the resulting outputs. In MCIL-MBT, we develop a formal model, verify the
formal model with model-checking, and conduct model-based testing in order
to remove the ambiguities as well as exhaustively cover variations of
execution sequences.

The remainder of our paper is organized as follows. In Section 2, we
present related work. In Section 3, we present our MCIL-MBT approach,
which, as noted above, has already been used by DENSO and eSOL and applied
to commercial OS implementations developed by eSOL. In Section 4, we
describe our results, with a discussion provided in Section 5. Finally, in
Section 6, we offer our conclusions and avenues for future work.
\section{Related Work}
For related work, we start with a joint project spanning from 2006 to 2015
that was conducted by JAIST with DENSO and Renesas Electronics organizations to
verify a commercial OSEK/VDX-compliant automotive OS developed by Renesas
Electronics using formal conventional methods \cite{JAIST5}. This OS was
already installed in many vehicles, and for this joint project, a design
model of the OS was developed and verified. Here, testing was conducted to
ensure that the OS implementation conformed to the given design model
\cite{JAIST2}. Verification of the design model was achieved through
theorem proving \cite{JAIST4} and satisfiability-modulo-theories (SMT)
solvers \cite{JAIST1,JAIST3}, as well as model-checking. Unlike our black
box strategy, these approaches used a white box approach.

More generally, formal OS verification has been studied extensively, as
widely surveyed by Klein \cite{OSV}. The most notable study in recent
years is the formal verification of an L4 kernel \cite{L4}, with the
verified kernel released as the seL4 kernel \cite{L4SE,L4SE2}. The
specifications of the L4 kernel are described in Haskell, with its
functional correctness, safety, and security proven by Isabelle/HOL
\cite{ISABELLE}. An instruction-level formal description of an ARM
processor, which is called ARM in HOL \cite{ARMINHOL}, was used in the
verification. The approach taken here was to formally prove the
correctness of the kernel implementation via theorem proving. We instead
make use of model-checking and MBT, though these approaches should be
complementary to one another.

Embedded OS implementations are specialized to support concurrent
multi-task or multi-thread scheduling since real-time properties are
crucial to embedded systems. In \cite{DEOS}, Penix {\it et al.} verified
an embedded OS scheduler called DEOS by using a model-checking tool called
SPIN \cite{SPIN}, which was developed for aerospace systems by
Honeywell. There are a number of studies, as summarized below, that
focused on verifying OSEK/VDX-based OS implementations. In
\cite{YUNJA1,YUNJA2,OSEK1}, Choi {\it et al.} verified the safety of the
source code for Trampoline \cite{TRAMPOLINE}, an open source OS, by
applying model-checking tools SPIN and SMV \cite{SMV}. Huang {\it et al.}
\cite{ECNU1} and Shin {\it et al.} \cite{ECNU2} verified a commercial OS
called ORIENTAIS, which was developed by iSoft using a process algebra
\cite{CSP}, a model-checking tool \cite{PAT}, and an automatic theorem
proving-based verification tool \cite{VCC}. Tigori {\it et al.}
\cite{OSEK1} proposed a method to generate a minimum amount of source code
that contains the functionality required to run a target application based
on a model-checking tool \cite{UPPAAL}. The above works applied a white
box approach that analyzes the source code of the given OS, develops a
model by reverse-engineering the code, and then verifies the model. Unlike
the above, our method adopts a black box approach in which we develop
formal specifications of the OS from the perspective of a user, i.e., a
company that accepts and verifies the OS against given
specifications. Further, we not only verify the model, but also conduct
exhaustive model-based testing by executing the actual OS.

Finally, Godefroid {\it et al.} proposed a tool called SAGE \cite{SAGE}
and its platform, SAGAN \cite{SAGAN}, to apply fuzz testing to OS
implementations. SAGE analyzes the source code of the OS and randomly
generates a large number of test cases, with the OS tested on a daily
basis. The progress and results of this testing are monitored and managed
by SAGAN. Since it is difficult to obtain accurate expected results,
simple properties such as system crashes are checked as part of the fuzz
testing. Unlike fuzz testing, our method focuses on scheduling-related
properties, such as which task is active and what the corresponding task
states are, all of which provides a more accurate picture than obtained
via fuzz testing.
\section{Model-Checking in the Loop with Model-Based Testing}
\subsection{Overview}
MCIL-MBT consists of (1) model development, (2) review and model-checking,
and (3) MBT. In (1), we develop a model for a target OS that both OS
developers and OS users can agree upon. Next, in (2), we conduct a review
and model-checking of the model from (1) to confirm that it works as
expected. In (3), we then confirm that the behavior of an actual OS
implementation is the same as the model by applying MBT. If we detect a
flaw during our MBT, the flaw is fixed and we iterate over these steps
until all test cases generated by the MBT are passed.
%
%
%
\subsection{Development of an OS Model}
\begin{figure}
  \begin{center}
  \centerline{\includegraphics[scale=0.6]{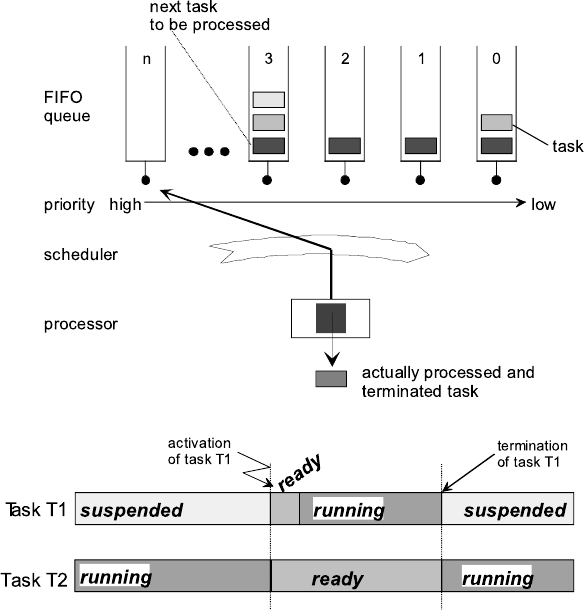}}
  \end{center}
  \caption{OS Specifications}
  \label{fig1}
\end{figure}

%
%
%
Scheduling details for each of AUTOSAR OS implementations are specified in
each specification document. Figure \ref{fig1} shows a portion of the
Classic AUTOSAR OS specifications document\cite{OSEK}. In the figure, the
upper side shows that there is a queue for each task priority level, while
the lower side shows the state changes that occur for two tasks T1 and T2
as time progresses left-to-right in the diagram. How these states are
changed is determined based on the queues shown in the upper side of the
figure. If the priority of T1 is higher than that of T2 and they are fully
preemptive, if T1 is activated while T2 is running, T1 will change to the
running state while T2 will change to the ready state. If instead two
tasks with the same priority are activated, the first activated task
enters the running state, as determined by the queues. As is evident from
the examples shown, the scheduling of tasks in all cases is specified as a
computation based on the priorities and queues; however, the document does
not define scheduling results in all cases.

From the above, we note that it is certainly difficult to describe
scheduling decisions in all cases in a specification document. It is far
preferable to represent the scheduling algorithm in an executable
form. Therefore, in MCIL-MBT, we specify scheduling as an executable
description using PROMELA, a specification language for the model-checking
tool SPIN. More specifically, the PROMELA description provides a formal
specification of an OS, and this formal specification allows us to remove
ambiguities at a very early phase of development, thereby avoiding flaws
introduced by such ambiguities in succeeding phases \cite{MYTH}. In
MCIL-MBT, creating the formal specification greatly reduces the risk of
backtracking in the development process caused by the discovery of flaws
at a later phase of development---especially at the end-phase in which
users accept the OS. In this paper, we refer to this formal specification
as the {\it OS model}. The behaviors defined by the APIs that make up the
OS are described in the OS model. For example, the behavior of
ActivateTask, which is one of Classic AUTOSAR OS APIs, is described as
calculating the state changes of tasks by enqueueing a task ID provided as
an argument to a queue according to its priority.
%
\subsection{Review and Model-Checking}
The quality of an OS model must be as high as possible since the OS plays
a central role in MCIL-MBT. The OS model determines the state changes of
tasks caused by OS API calls with priorities and queues. As
such, this determination is complex, and it can be difficult to achieve a
high level of confidence in the correctness of the model by only reviewing
the OS model. Therefore, we construct state models that are simpler than
the OS model; we call these simpler models {\it use case models}.

\begin{figure}
  \begin{center}
  \centerline{\includegraphics[scale=0.35]{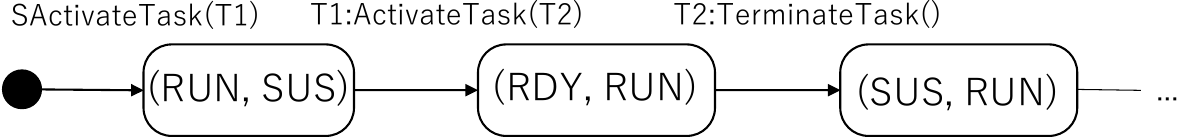}}
  \end{center}
  \caption{Invocation Sequence and Expected Results}
  \label{fig2}
\end{figure}

Figure \ref{fig2} shows an example of these use case models in which there
are two tasks T1 and T2. Here, T2 has a higher priority than T1, T1 is
preemptive, and both tasks are executed on the same core. Initially, T1 is
activated via the SActivateTask(T1) API call and subsequently
executed. Next, T2 is executed when T2 is activated by T1, i.e., we have
T1:ActivateTask(T2). Finally, T1 resumes its execution when T2 terminates,
as initiated by T2:TerminateTask(). The expected state changes of the
tasks are also shown in the figure as nodes with SUS, RUN, and RDY
indicating task states. Here, SUS indicates a suspended state in which a
task is not active, RUN indicates a running state in which a task is being
executed, and RDY indicates a ready state in which a task is waiting for
execution. In the figure, the first element of each pair represents the
expected state of T1, while the second element represents the expected
state of T2. As an example, the first two states and a transition between
them mean that the states of T1 and T2 are expected to be RDY and RUN if
T2 is activated by T1 when the states of T1 and T2 are RUN and SUS,
respectively. Such a use case model is easier to review because it is
focused on a smaller subset of functionality.
In our approach, SPIN is used to verify that an OS model conforms to
defined use case models after both users and developers agree with the
models in the review. The constructed use case models can be converted
into PROMELA descriptions, with the expected state changes described as
assertions, thus making it possible to check whether the behavior of the
OS model is as expected, i.e., as described in the use case models.

Use case models are more reliable than the OS model quite simply
because the former is simpler than the latter. Unfortunately, only limited
typical cases and expected results can be described via use case models as
it is impossible for them to cover all possible cases and describe all
possible behaviors included in the OS model. In our approach, we review
both the OS model and the corresponding use case models. In the review of
the OS model, we check the computation mechanisms of the scheduling, i.e.,
how state changes are determined. In the review of the use case models, we
check the typical behavior of the OS. Combining these two macro- and
micro-level reviews and conducting model-checking for each can practically
improve the reliability of the OS model.
\subsection{Model-Based Testing}
Testing the scheduling behavior of an OS is challenging because we must
execute the OS using all possible execution sequences produced by any
combination of OS API invocations. And beyond identifying all such
combinations, we must also describe expected results, i.e., a test
oracle. To test such complex and non-deterministic behavior of the given
OS, we adopt MBT to automatically generate an exhaustive set of test cases
and test programs from the given OS model by exploring its states using a
model-checking algorithm. The test oracle can also be obtained from the OS
model. Therefore, we check whether the behavior of the OS matches the OS
model or not.
\begin{figure}
  \begin{center}
  \centerline{\includegraphics[scale=0.45]{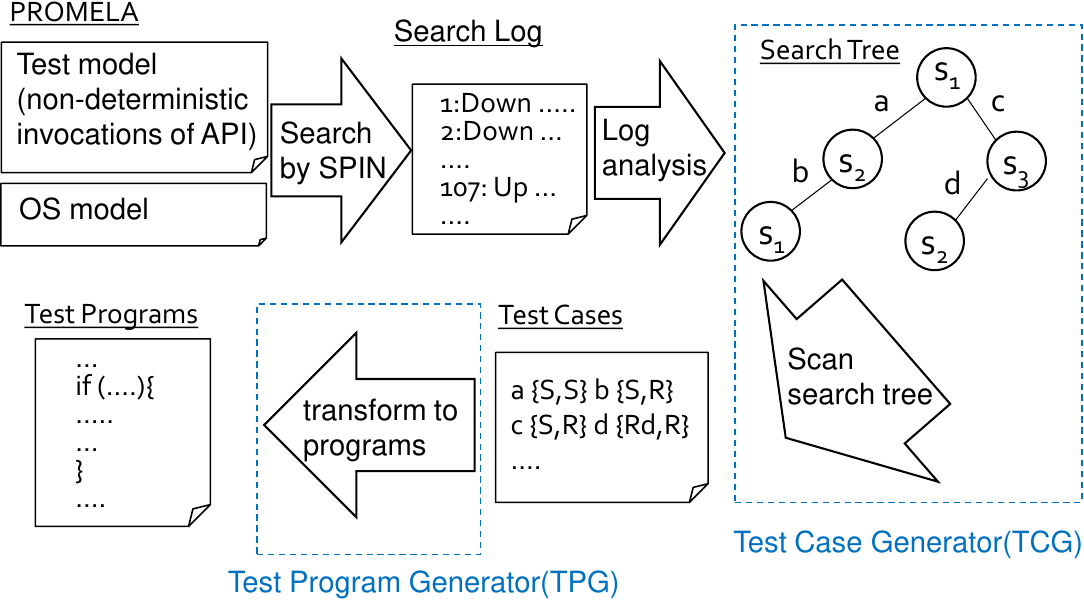}}
  \end{center}
  \caption{Automatic Generation of Test Cases}
  \label{fig4}
\end{figure}
In Figure \ref{fig4}, we describe a procedure for automatically generating
the test cases and the test oracle. The OS APIs described in the OS model are
non-deterministically invoked to explore the reachable states of the OS
model. We call a model describing such invocations a {\it test model}, and
we use SPIN to explore the reachable states. Here, SPIN has a function
that will output the status of the state search as a log, so by analyzing
this log, we create a search tree of the reachable states. Further, the
expected results of the API invocations can be obtained from the OS
model. As the OS model determines the state changes of the given tasks, we
can acquire the corresponding states that represent the expected results
of the API invocations at any point in the state search. Therefore, we
output such states to the log and create a search tree that includes
them. By traversing this search tree, we can obtain API invocation
sequences that visit all states of the OS model with the expected
results. These invocation sequences and expected results are simply called
our {\it test cases}, while the tool that automatically generates these
test cases is our {\it Test Case Generator (TCG)}.
The idea to generate test cases are basically the same as the one proposed
in \cite{JAIST5} though some extensions were needed for adapting it to AUTOSAR
OS. Due to space limitation, we omit their details in this paper.

From the above test cases, we generate test programs via a tool called the
{\it Test Program  Generator (TPG)}. Invoking APIs and checking expected
results must be performed on an actual running task because each of the
test programs is an application that runs on an OS
implementation. Therefore, it is necessary to allocate test cases to each
task; however, the test cases generated by TCG do not take such allocation
into account. Therefore, in TPG, test cases are analyzed such that they
can be realized as applications.

Since a large number of test programs is generated, it is essential to
execute the generated applications in parallel. Therefore, we make use of
a large-scale computer cluster called StarBED \cite{STARBED} to execute
these test programs in parallel. Each expected result is checked via an
assertion in the corresponding test program, and each assertion violation
is recorded in a log. Test results and logs are stored in an external
storage system and analyzed offline after testing is complete.

Model-checking verifies a model developed manually and usually differs
from an actual OS implementation. Therefore, a gap exists between the
model and the actual implementation, which in turn causes a problem in
that a property verified in the model might not be preserved in the
implementation. To counter this problem, in MCIL-MBT, we use an OS model
verified by model-checking to automatically generate test programs that
compare the behavior of the OS model with an actual OS
implementation. This combination of model-checking and testing enables us
to avoid the gap noted above and seamlessly connect the model and the
implementation.
\subsection{OS Acceptance Testing}
In the development of an automotive system, concurrent tasks are designed
according to their time constraints and importance. An accurate
understanding of the scheduling capabilities provided by an OS is
therefore particularly important. Further, this understanding must be the
same for both an OS developer who implements an OS and a user who makes
use of that OS. The user expects that the OS implemented by the OS
developer correctly realizes the scheduling algorithm that the user had in
mind when the OS was accepted. Given this, two types of flaws can be
detected here, i.e., either a bug in the implementation or a discrepancy
between user and developer expectations---in other words, how tasks are
scheduled is interpreted differently by the user and the developer.

OS acceptance testing is usually conducted when an implemented OS is ready
for users to perform their testing. It is difficult to exhaustively test
the behavior of an OS during acceptance testing because the user may only
test via black box testing; further, there is an excessively large number
of behavioral variations within even a rudimentary OS
implementation. Therefore, in our MCIL-MBT approach, the behavior of the
OS is agreed upon in advance by users and developers based on a formally
described OS model. Next, we use MBT to exhaustively check whether the OS
implementation conforms to the agreed upon OS model or not.

If a flaw is found during the MBT, the flaw is handled according to its
type. If caused by a bug in the OS implementation, the MBT is conducted
again after fixing this bug. Otherwise, if the bug is caused by a
discrepancy in what the developer and the what the user expects in terms
of behavior, the OS model and use case models are modified after coming to
a consensus. Next, model-checking and MBT are conducted again based on the
modified OS model and use case models. The quality of the OS
implementation is improved as a result of fixing flaws and resolving
discrepancies via the iterative approach of modifying the OS model,
performing model-checking, and conducting MBT. The OS implementation is
finally accepted after all MBT test cases pass.
%
\section{MCIL-MBT Results}
\subsection{Organization}
We applied MCIL-MBT in conjunction with DENSO and eSOL, which represent
the user and developer sides, respectively. DENSO primarily developed OS
models and corresponding use case models based on AUTOSAR OS
specifications, then reviewed and verified them. DENSO contacted eSOL to
determine the exact behavior of OS implementations in developing the OS
models if unknown behaviors were encountered. DENSO CREATE dispatched
engineers to develop and review the OS models and use case models whenever
additional human resources were necessary. We applied our MCIL-MBT
approach to eMCOS OS implementations \cite{EMCOS}, which are commercial
automotive OS implementations developed by eSOL. As there are various
specifications for the eMCOS OS implementations, in this paper, we focus
on two of them, i.e., eMCOS AUTOSAR and eMCOS POSIX, which are Classic
AUTOSAR OS and a POSIX OS, respectively. 
\begin{figure*}[t]
  \begin{center}
  \centerline{\includegraphics[scale=0.45]{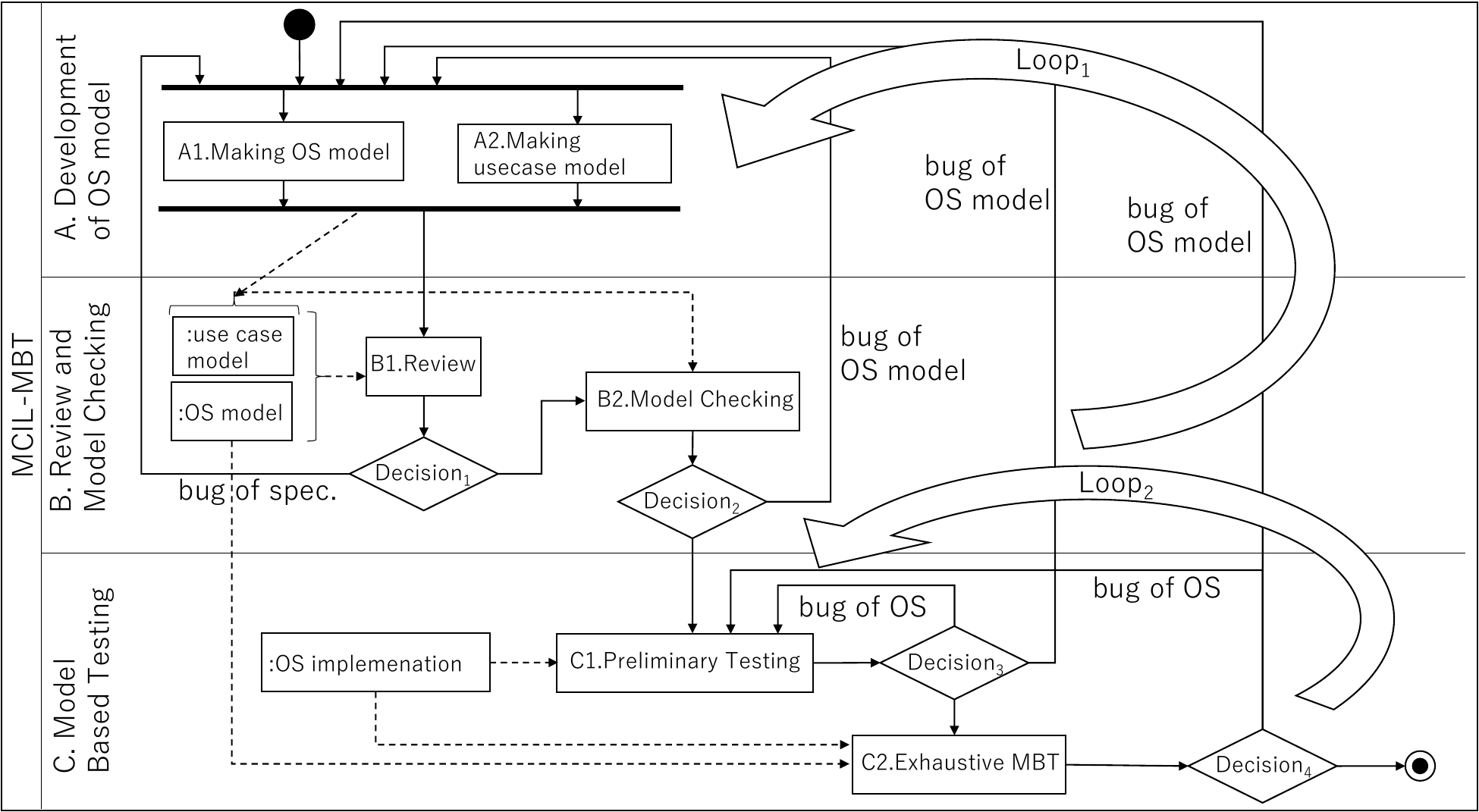}}
  \end{center}
  \caption{MCIL-MBT Workflow}
  \label{workflow}
\end{figure*}
\subsection{Workflow}
Figure \ref{workflow} shows the workflow of our MCIL-MBT approach. We
divide this workflow into the following three key phases: (A) an OS model
development phase; (B) a review and model-checking phase; and (C) a MBT
phase. Corresponding to the figure, in phase (A), an OS model (A1) and use
case models (A2) are developed in parallel. In phase (B), the OS model and
use case models are reviewed (B1). Note that we backtrack to phase (A) if
we find a bug in the OS model of in the understanding of the OS
specifications ($Decision_1$); in this case, the review is conducted again
after the OS model and use case models are modified to address the
discrepancy or fix the bug ($Loop_1$). If this review is passed, we
perform model-checking (B2). And similar to the above, we backtrack to
phase (A) and re-execute phase (B) if a flaw is found during
model-checking ($Loop_1$).

Once we are confident of the correctness of the OS model ($Decision_2$),
we perform preliminary testing (C1) in which we test an OS implementation
using a small number of fairly typical test cases. This is to detect
simple flaws in advance since the full execution of MBT takes a
substantial amount of time. As noted above, there are two types of flaws
we again look for at this stage ($Decision_3$), namely a bug in the OS
implementation or a discrepancy in the understanding of the OS
specifications. If the former, we execute (C1) again after fixing the flaw
in the OS ($Loop_2$); and if the latter, we backtrack to phase (A), modify
the OS model and use case models, then perform phases (B) and (C1) again
($Loop_1$).

Moving forward from here, we perform MBT if the preliminary testing stage
passed ($Decision_3$). And once again, there are two types of flaws that
we might encounter here ($Decision_4$); as with $Decision_3$, we have
either a bug in the OS implementation or a discrepancy in the
understanding of the OS specifications. These are treated in the same way
as shown for $Decision_3$. In the former, we execute (C1) and (C2) again
after fixing the flaw in the OS ($Loop_2$), while in the latter, we
backtrack to phase (A), modify the OS model and use case models, then
perform phases (B), (C1), and (C2) again ($Loop_1$). Finally, MCIL-MBT
terminates successfully if all test cases are passed in (C2).

There are two loops shown in the workflow, i.e., $Loop_1$ and
$Loop_2$. Here, $Loop_1$ goes back to phase (A) if we detect a flaw in the
OS model due to a discrepancy in the understanding of the OS
specifications. For $Loop_2$, we perform preliminary testing and
exhaustive MBT again if a bug in the OS implementation is detected. These
critical looping mechanisms motivated us to come up with the {\it
  Model-Checking in the Loop} portion of MCIL-MBT, i.e., we perform
model-checking to verify the OS model and employ MBT, which uses a
model-checking algorithm for testing the OS implementation.
\begin{table*}[htb]
\caption{MBT Results}
\label{result}
\begin{center}
  \begin{tabular}{l l l l l l l l}\hline
    target & model & tc & total time & node time & nodes (sim procs) & flaws(OS) &
    flaws(spec)\\ \hline
    Classic(S,B) & 1755 & 185741 & 5156413s ($\fallingdotseq$59.7d) &
    14378.03s ($\fallingdotseq$4h) & 29 (20)
    & 0 & 1\\
    Classic(S,A) & 2134 & 135081 & 4349349s ($\fallingdotseq$50.3d) &
    6795.59s ($\fallingdotseq$1.9h) & 54 (20)
    & 0 & 0 \\
    Classic(M,B) & 1879 & 134756 & 1464166s ($\fallingdotseq$16.94d) &
    2100.61s ($\fallingdotseq$35m) & 54 (20)
    & 0 & 0 \\
    Classic(M,A) & 2183 & 117263 & 1428272s ($\fallingdotseq$16.53d) & 2073.23 ($\fallingdotseq$0.57h) &
    54 (20) & 0 & 0 \\
    Classic(M,P) & 2840 & 625863 & 27360149s ($\fallingdotseq$216.66d) & 28825.96 ($\fallingdotseq$8h) &
    84 (20) & 2 & 0 \\
    Adaptive(M,B) & 3941 & 183537 & 98029724 ($\fallingdotseq$1134.60d) & 93548s
    ($\fallingdotseq$1d1h59m8s) & 52 (23) & 4 & 4 \\
    Adaptive(M,N) & 4102 & 261571 & 221703978s ($\fallingdotseq$2566.02d) &
    226317s($\fallingdotseq$2d14h51m57s) & 52 (23) & 5 & 1 \\ \hline
  \end{tabular}
\end{center}
\end{table*}
\subsection{Results}
\subsubsection{OS Models}
Our MBT results are summarized in Table \ref{result}. We divided the
functionality of a target OS into multiple features, then created an OS
model and performed MBT for each of these target OS implementations in an
effort to reduce the number of test cases. We created five OS models for
eMCOS AUTOSAR, which we represent in the table as Classic(S,B),
Classic(S,A), Classic(M,B), Classic(M,A), and Classic(M,P). Here, the
eMCOS OS can be configured according to a target platform. The
Classic(S,B) and Classic(S,A) rows show results with a configuration in
which eMCOS AUTOSAR works as a single-core OS. The former focuses on basic
(B) scheduling functions, whereas the latter focuses on alarm (A)
functions. The Classic(M,B), Classic(M,A), and Classic(M,P) rows represent
results with a configuration in which eMCOS AUTOSAR works as a multi-core
OS focused on basic (B) scheduling functions, alarm (A) functions, and
spinlock (P) functions, respectively. Note that MBT results are shown for
Classic(S,B), Classic(S,A), and Classic(M,P); however, other data such as
the details of the flaws detected were unfortunately lost due to failures
in our external storage system. We also created two OS models for eMCOS
POSIX, represented as Adaptive(M,B) and Adaptive(M,N), which focused on
basic (B) scheduling functions of Adaptive AUTOSAR OS (AMP version) and
eMCOS POSIX native (N) functions, respectively. We did not focus on the
SMP feature of eMCOS POSIX this time.  AUTOSAR OSs are highly configurable
with respect to the scheduling. We did not focus on all of possible
configurations but the ones that user sides, that is, DENSO use in their
developments.

\subsubsection{Review and Model-Checking Results}
An OS model was developed for each of the OS features. In Table
\ref{result}, the {\it model} column shows the number of lines for each OS
model. We also developed use case models for each of the OS features, as
detailed in Table \ref{usecase}; the {\it num}, {\it api}, and {\it cost}
columns show the number of use case models developed, the average number
of API invocations in each use case model, and the cost in person-months
required to develop and review the use case models, respectively.  The use
cases were constructed in the following two steps. Firstly, we made a use
case model for each functionality of AUTOSAR OS. Then, we analyzed
interactions of multiple functionalities. If the interaction affects the
scheduling, we made another use case model for it.

We described the use case models using state machine diagrams created via
a UML drawing tool called PlantUML \cite{PLANTUML}. These use case models
were automatically converted into PROMELA descriptions and combined with
the OS models for automated model-checking via SPIN. In Table
\ref{usecase}, the {\it line} and {\it time} columns show the average
number of lines of the PROMELA descriptions and the time (in minutes)
required to complete the model-checking, respectively. Note that the PC
used for model-checking was a 64-bit Windows 10 Enterprise machine with an
Intel Core i7-7700 3.60GHz CPU and 16 GB of memory.

We regularly reviewed the OS models and use case models. Further, we
verified that the behavior of the OS models conformed to the use case
models by feeding them into SPIN after manually reviewing them. Table
\ref{modelbug} describes the number of flaws found in the review and
model-checking steps for the Adaptive(M,B) and Adaptive(M,N) models. Here,
the {\it Review} row shows the number of flaws found in the review, while
the {\it MC} row shows the number of flaws found in model-checking. The
{\it flaw(UC)} and {\it flaw(OS)} columns show the number of flaws found
in the use case models and OS models, respectively. Likewise, the {\it
  spec(UC)} and {\it spec(OS)} columns show the number of discrepancies in
the understanding of the OS specifications found in the use case models
and OS models, respectively. Note that the Adaptive(M,N) OS model was
developed by modifying and extending the Adaptive(M,B) model. We observe
that the number of detected flaws in the Adaptive(M,N) model was small
since the already verified Adaptive(M,B) model was reused to develop
it. Further, the number of flaws found in the use case models (i.e., {\it
  flaw(UC)}) in the Adaptive(M,N) model was relatively large because a new
member of the team was assigned to develop these use case models and was
initially unfamiliar with them.
\begin{table}
  \caption{Use Case Models}
  \label{usecase}
  \begin{center}
    \begin{tabular}{l l l l l l}\hline
      & num & api & cost & line & time (min) \\ \hline
      Adaptive(M,B) & 498 & 6.2 & 3.7 & 75.7  & approx. 5 \\
      Adaptive(M,N) & 496 & 6.4 & 1.25 & 79.1  & approx. 10 \\ \hline
    \end{tabular}
  \end{center}
\end{table}
\begin{table}
  \caption{Review and Model-Checking Results}
  \label{modelbug}
  \begin{center}
    \begin{tabular}{l l l l l}
      \multicolumn{5}{l}{Adaptive(M,B)} \\ \hline
      & flaw(UC) & flaw(OS) & spec(UC) & spec(OS) \\ \hline
      Review & 5 & 37 & 0 &  1 \\
      MC & 0 & 1 & 1 & 0 \\ \hline
      \multicolumn{5}{l}{} \\ 
      \multicolumn{5}{l}{Adaptive(M,N)} \\ \hline
      Review & 5 & 0 & 2 &  0 \\
      MC & 0 & 0 & 0 & 0 \\ \hline
    \end{tabular}
  \end{center}
\end{table}
\subsubsection{MBT Results}
As noted previously, for our MBT, we used a large-scale computer cluster
called StarBED that consisted of more than 1000 PCs. We developed a
mechanism to conduct our MBT in parallel and store results in an external
storage system after installing a simulator of a target CPU on top of a
CentOS instance on each PC. Preparing StarBED did not require much time,
only one day.

Back to Table \ref{result}, the {\it tc} column shows the number of test cases generated from an OS model. Further, the {\it nodes} column shows the number of nodes (i.e., PCs) used to perform MBT. Note that each node was a Dell PowerEdge C6220 with a dual Intel Xeon E5-2650 processor (2.00GHz/8 core) and 128 GB of memory. Given this configuration, MBT was conducted in parallel. Multiple simulator processes were executed on each node with test programs run on each simulator process. In the table, the {\it nodes (sim procs)} column shows the number of nodes used in the execution of the test programs, with the number of simulator processes executed on each node shown in parentheses. The {\it total time} column shows the time taken to run all test programs, which is essentially the sum of CPU times on all of the PC nodes; in other words, this column represents how long it takes to run all test programs if they are executed sequentially. The {\it node time} column shows the average time spent in each node, which essentially represents the actual time required to complete our MBT. Finally, the {\it flaw(OS)} and {\it flaw(spec)} columns show the numbers of flaws detected in our MBT, representing bugs in the OS implementation itself and discrepancies in the understanding of the OS specifications, respectively.

Finally, in Table \ref{bug}, we present details regarding the detected
flaws. Flaws are shown by the {\it type} column to be either OS or Spec
flaws representing either bugs in the OS implementation or discrepancies
in the understanding of the OS specifications, respectively. We also show
the minimum ({\it min}), maximum ({\it max}), and average ({\it ave})
number of steps required to find each of the flaws since multiple test
cases failed in order for us to identify a single flaw during our
MBT. Note that only the minimum number of steps are shown for the Spec(2)
through Spec(4) rows because these flaws were identified during
preliminary testing. The {\it api} and {\it rc} columns show the API in
which each flaw occurred and the specific factor that caused it,
respectively. For the {\it rc} column, ercd, priority, and run represent
flaws in error codes, priority changes, and execution control of tasks
such as blocking/unblocking a task, respectively. 
\begin{table}[h]
  \caption{Flaws Encountered}
  \label{bug}
  \begin{center}
    \begin{tabular}{l l l l l l}
      \multicolumn{6}{l}{Adaptive(M,B)} \\ \hline
      type & min & max & ave & api & rc\\ \hline
      OS(1) & 4 & 7 & 4.4 &  mutex\_trylock & ercd\\
      OS(2) & 25 & 28 & 25.009 & mutex\_lock & priority\\
      OS(3) & 11 & 34 & 24.7  & mutex\_lock & priority \\
      OS(4) & 11 & 34 & 24.7  & mutex\_unlock & run\\
      OS(5) & 25 & 36 & 28.0 & mutex\_timedlock & priority \\
      Spec(1) & 47 & 81 & 61.4 & mutex\_lock & ercd \\
      Spec(2) & 9 & NA & NA & cond\_wait & run \\
      Spec(3) & 3 & NA & NA & cond\_signal & run \\
      Spec(4) & 3 & NA & NA & setschedprio & priority \\ \hline
      \multicolumn{6}{l}{} \\
      \multicolumn{6}{l}{Adaptive(M,N)} \\ \hline
      OS(6) & 3 & NA & NA &  mutex\_lock & ercd\\
      OS(7) & 8 & NA & NA & mutex\_unlock & priority\\
      OS(8) & 11 & NA & NA  & mutex\_lock & run \\
      Spec(5) & 22 & 45 & 37.6 & mutex\_unlock & run\\ \hline
    \end{tabular}
  \end{center}
\end{table}

\subsection{Iterations}
As noted previously, MCIL-MBT is an iterative approach. More specifically,
we iterate over $Loop_1$ to modify the OS models and use case models,
where as we iterate over $Loop_2$ to modify the OS implementation. In
Table \ref{iterations}, we summarize the number and time periods of such
iterations. Since we did not initially record precise data while
conducting MCIL-MBT, we determined these results by interviewing engineers
who participated in the practice about their specific experiences. Note
that in the table, the $Decision_1$ through $Decision_4$ points represent
decision points shown in Figure \ref{workflow} above. Further, each
iteration is represented as a $Decision_i\mbox{-}Loop_j$ pairs, where
$1\leq i\leq 4$ and $j=1,2$; each pair indicates that $Loop_j$ occurred in
$Decision_i$. Note that $Loop_1$ occurred in $Decision_i, where 1\leq i
\leq 4$, while $Loop_2$ occurred only in $Decision_3$ and
$Decision_4$. The {\it iterations} and {\it period} columns show the
number of iterations and the time period (in days) required to modify the
OS models and use case models, respectively. Here, the time period is not
shown for $Loop_2$ because it depended on the outside of MCIL-MBT, that
is, the timing to modify the OS implementation by eSOL. 

\begin{table}
  \caption{Iteration Results}
  \label{iterations}
  \begin{center}
    \begin{tabular}{l l l}\hline
      point & iterations & period \\ \hline
      $Decision_1$-$Loop_1$ & 1 - 3 & 1 - 7 day(s) \\
      $Decision_2$-$Loop_1$ & 1 - 3 & 1 - 7  day(s) \\
      $Decision_3$-$Loop_1$ & 1 - 3 & 5 - 10 days \\
      $Decision_3$-$Loop_2$ & 1 - 3 & NA \\
      $Decision_4$-$Loop_1$ & 2 - 5 & 5 - 10 days \\
      $Decision_4$-$Loop_2$ & 2 - 5 & NA \\ \hline
    \end{tabular}
  \end{center}
\end{table}
\subsection{Examples of Detected Flaws}
In this section, we share a few example flaws that we detected via our MCIL-MBT approach. First, we show an example of a discrepancy in the understanding of the OS specifications; in this example, the API call is ChainTask(), an event mask in eMCOS AUTOSAR. If ChainTask is invoked with an argument that designates the next task to be dispatched, the task which invokes it is terminated, and the designated task changes to a ready state. For example, given tasks T1 and T2 in the running and suspended states, respectively, an invocation of ChainTask(T2) changes T1 to a suspended state and T2 to a ready state. The argument to ChainTask() can be the same as the task that invokes ChainTask(), i.e., ChainTask(T1) can be invoked by T1. The behavior of ChainTask() is described in the Classic AUTOSAR OS (OSEK OS) documentation as follows:
\begin{itemize}
\item If the succeeding task is identical with the current task, this does not result in multiple requests. The task is not transferred to the suspended state, but will immediately become ready again.
\item When an extended task is transferred from suspended state into ready state all its events are cleared.
\end{itemize}
The understanding by the users was that the event would not be cleared if task T1 invokes ChainTask(T1) because T1 does not enter the suspended state. Conversely, the understanding by the developers was that the event would be cleared in this case because it should work the same as the reactivation of T1, i.e., the state of T1 is changed from suspended to ready after it is changed from ready to suspended.

\begin{figure}
  \begin{center}
  \centerline{\includegraphics[scale=0.4]{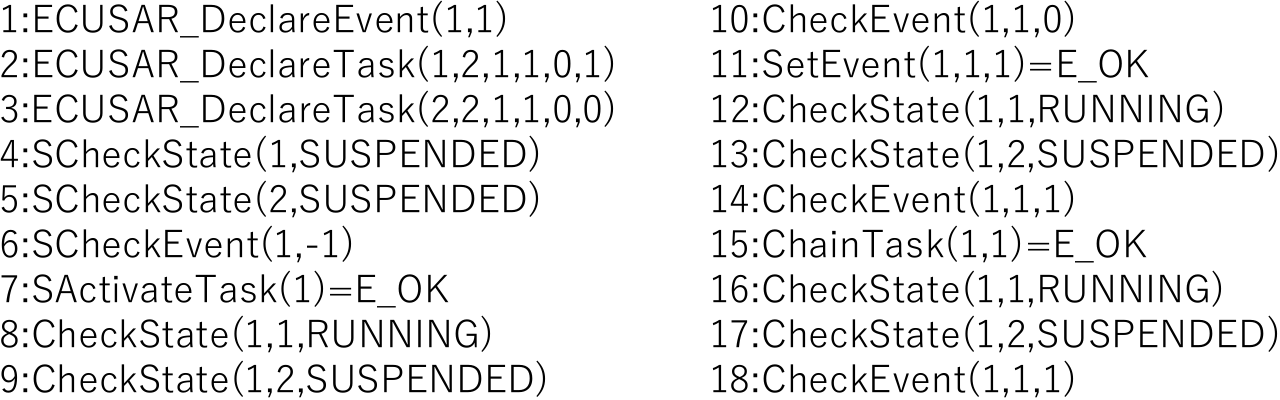}}
  \end{center}
  \caption{Generated Test Case that Detected a Flaw}
  \label{fig6}
\end{figure}

This flaw was detected in the test case shown in Figure \ref{fig6}. In the figure, API invocations are preceded with a sequential number that represents the order in which each step is invoked. Steps 1 through 3 declare an event and two tasks. These tasks have IDs 1 and 2, and we refer to these tasks as T1 and T2 here, respectively. T1 and T2 run on the same core, both with the same priority of 2. Steps 7 through 10 represent startup procedures. More specifically, in Step 7 specifies that T1 is activated with an expected return value of \verb|E_OK|, which is a return code that indicates a successful activation. In Steps 8 and 9, \verb|CheckState(x,y,z)| means that task \verb|x| checks whether the state of task \verb|y| is \verb|z| or not. In Step 10, \verb|CheckEvent(x,y,z)| means that task \verb|x| checks whether the mask of event \verb|y| is \verb|z| or not. Similarly, in Steps 11 and 15, \verb|SetEvent| and \verb|ChainTask| are invoked, with their expected return values checked in Steps 12 through 14 and Steps 16 through 18. The checking of the various expected values are converted into assertions in the automatically generated test programs. Here, an assertion violation is reported at Step 18 when a test program converted from the test case is executed.
Finally, we present an example implementation bug involving the acquisition of a spinlock in eMCOS AUTOSAR. In the Classic AUTOSAR OS, if multiple spinlock resources exist, the order in which they are acquired within the same core is specified as part of the configuration, a setting that has to be maintained at runtime. Here, error code \verb|E_OS_NESTING_DEADLOCK| is returned if the order is violated at runtime. We detected a case in which this error code was not returned even though the order was indeed violated. We omit the details of the test case because it was too long to show in this paper---it consisted of 37 distinct steps.
\section{Discussion}
\subsection{Flaw Detection}
Applying our MCIL-MBT approach, 17 flaws were detected even though the target OS implementations were already verified by the OS developers in advance. We conclude that by applying MCIL-MBT, we were able to detect flaws that were difficult to detect via conventional methods typically used in industry. In the subsections that follow, we discuss detected flaws from two distinct points of view, i.e., flaws caused by discrepancies in the understanding of the OS specifications and flaws caused by bugs in the OS implementation.
\subsubsection{Discrepancies in the understanding of the OS specifications}
As shown in Table \ref{bug} above, most of the flaws caused by
discrepancies in the understanding of the OS specifications were detected
at early phases, including the development of the OS models and
preliminary testing. We therefore infer that by writing formal
specifications of the OS, these OS models worked effectively in detecting
flaws; however, there were exceptions, i.e., flaws Spec(1) and Spec(5)
shown in Table \ref{bug} above were detected later in our MBT.
In Table \ref{iterations}, the number of iterations caused by $Decision_4$
was greater than the number of iterations caused by other decisions.
These observations imply that flaws caused by discrepancies in the
understanding of the OS specifications can be missed if only typical
sequences of API invocations are considered. Further, it is difficult to
detect flaws using conventional methods here since the number of steps in
the corresponding test cases is quite long.

The DENSO/DENSO CREATE sides asked the eSOL side about unknown behavior in
the OS models, then modified the OS models such that they agreed with one
another. Although it is desirable that both sides develop and review the
OS models, it is difficult from a practical point of view to successfully
collaborate on those activities since developers and users belong to
different companies and have different development styles. Applying our
MCIL-MBT approach made it possible to more readily detect discrepancies.

As noted previously, only typical execution examples are described in the
AUTOSAR OS specifications. In fact, only 10 execution examples are
described in the OSEK/VDX specifications. A document on conformance
testing titled MODISTARC \cite{MODISTARC} is provided for an
OSEK/VDX-compliant OS. In MODISTARC, there are 139 test cases that cover
invocations of OS APIs, with each test case defined as a a single API
invocation and its corresponding precondition. In DENSO, test cases were
created to cover typical API invocations in conventional acceptance
testing, and 95 test cases were used to test OS implementations for the
behaviors that we focused on in our MCIL-MBT approach. Here, it was
necessary to take invocation sequences of APIs into account in order to
accurately test the scheduling of an OS since scheduling depends on the
state of the OS, which varies according to the order in which API
invocations are made. Conversely, it is challenging to define all such API
invocation sequences. Therefore, in our MCIL-MBT approach, we did not
directly define the invocation sequences, letting the OS models generate
them instead. Such OS models allow us to share more accurate
specifications of the given OS between developers and users, thereby
further encouraging detection of any discrepancies in the understanding of
the OS specifications.

When engineers develop applications that run on an OS, they assume that
the OS works according to its specification. The quality of the
applications may decrease if their understanding of the specification is
ambiguous. Certainly OS models are useful for such developers to clearly
understand the expected behavior of the OS.
\subsubsection{Bugs in the OS implementation}
Applying our MCIL-MBT method successfully detected OS implementation bugs that were not already detected in advance. More specifically, use of our MCIL-MBT approach was able to detect such bugs since it generates test cases not typically covered by conventional testing methodologies. More specifically, our MBT took various combinations of API invocations based on OS models into account, generating corresponding invocation sequences. As shown in Table \ref{bug} above, the average number of steps per test case that successfully detected flaws was approximately 25, and the minimum number of steps in a test case was more than 11 except for OS(1), OS(6), and OS(7). Although it is very difficult to manually construct such test cases, our MCIL-MBT approach was able to uncover and detect these bugs.

In acceptance testing, we usually perform black box testing that inherently does not allow us to view or analyze the code of a given OS implementation. While we know it is theoretically impossible for black box testing to cover all states of a system we are testing \cite{ATCT} and therefore also impossible to cover all states of an OS implementation using our MCIL-MBT approach, MCIL-MBT covers all possible combinations of API invocations from the point of view of the OS specifications. Using such a detailed and specialized approach was very effective in detecting implementation bugs. In other words, even though testing was based solely on the OS specifications, our approach effectively detected implementation bugs not found using conventional methods.
\subsection{Ensuring Quality in the OS Models}
As another important conclusion from our work described above, we note that the quality of the OS models must be high because these models play such a crucial role in our MCIL-MBT method. Given that the OS models ranged from 1755 to 4102 lines, as shown in Table \ref{result} above, it is difficult to ensure a high level of quality only via a manual review process. Therefore, we adopted model-checking to verify the OS models. In our model-checking approach, expected properties must be described as temporal logic, but it is difficult for engineers to describe them exactly. Although a method for defining typical temporal logic formulas as patterns and then using these patterns in practice has been proposed in \cite{SPECP}, such patterns are not suitable for scheduling properties, because we need to check state changes step by step based on invocation sequences. Therefore, instead of using temporal logic formulas, we used observers that invoke the APIs described in the OS models and synchronously observe their subsequent state changes. This technique is known as using {\it versatile synchronous observers} and is described as being easy to use in practice \cite{SOBS}.

If the observers are too complex, it is difficult to have confidence in
their correctness. Therefore, we defined an observer as a simple use case
model in our MCIL-MBT approach. As shown in Table \ref{usecase} above, the
average number of API invocations in each use case model was approximately
six. It is easy for engineers to create and review these simpler use case
models, because the complexity here is equivalent to that of the examples
that appear in the AUTOSAR OS specifications. Further, although there are
a large number of use case models, reviewing them does not require
expertise in model-checking and could be distributed out to several
engineers.

As shown in Table \ref{modelbug} above, most flaws in the OS models and use case models were detected by the review process, with only a few flaws detected by model-checking. We therefore conclude that model-checking worked effectively as these were flaws that were not detected during the review. Further, the number of flaws detected in the review was smaller than that of the OS models despite the fact that the total number of lines in the use case models is much larger than that of the OS models. We conclude that the use case models are more reliable than the OS models given their relative scales. Therefore, we think that the quality of the OS models can be practically increased by verifying the OS models against the use case models via model-checking.
\subsection{Testing OS}
As shown in Table \ref{result} above, hundreds of thousands of test cases for each of the OS models were generated using our MCIL-MBT approach. This implies that a huge number of states must be taken into account when conducting OS implementation testing; further, it is practically impossible to define the accurate behavior of OS implementations based on their execution sequences. From our results, the total time taken to complete the testing ranged from weeks to months for the Classic AUTOSAR OS up to several years for the Adaptive AUTOSAR OS, if such testing was performed sequentially. As noted previously, we performed testing in parallel on the large-scale computer cluster StarBED since each test program can be executed independently. Without a doubt, StarBED allowed us to complete the testing much more quickly, i.e., in several hours for the Classic AUTOSAR OS and across a few days for the Adaptive AUTOSAR OS. As such, StarBED was a good and reasonable choice for the practical use of our MCIL-MBT approach.

As part of our present work, we needed to generate a test program from a test case such that the test program can invoke APIs in the same order defined in the test case. The AUTOSAR OS supports multi-cores, and tasks assigned to different cores are independently executed on these cores. In such multi-core OS implementations, simply ordering API invocations of tasks in the test program does not realize the order defined in the test case. To correctly realize it, we inserted {\it busy-waits} between the API invocations of the test program. We call such a test a {\it sequence test}, noting that sequence tests make it easier to detect logical flaws, such as specification violations, which are important for acceptance testing. More specifically here, we need to execute the same test program numerous times to invoke the APIs in a specific order without the use of busy-waits \cite{DSLMC}. Beyond the use of sequence tests, there are other ways to test OS implementations based on the OS model. For example, we can execute OS implementations freely and check whether results of the execution conform to the OS models or not. Currently, we are investigating other ways to test OS implementations based on the OS models.
\section{Conclusion}
In this paper, we proposed an integrated verification and testing methodology called MCIL-MBT for verifying automotive OS implementations. Our method consists of OS modeling, reviewing and model-checking of these created OS models, and MBT of specific OS implementations. For the MCIL portion of our approach, OS models are verified using model-checking, test cases are automatically generated using a model-checking algorithm, and these generated test cases are iterated over until OS implementations are accepted by users. Our MCIL-MBT approach was successfully applied by DENSO and eSOL and also applied to commercial OS implementations developed by eSOL. Although those OS implementations were verified by conventional methods, we succeeded in applying our MCIL-MBT approach to detect additional flaws not already detected in advance. DENSO, the user side, has primarily focused on conducting development and review of OS models and MBT of OS implementations. In the future, eSOL, the OS developer side, plans to apply MCIL-MBT before such OS implementations are released. We also plan to expand target OS implementations, in particular, the SMP version of the Adaptive AUTOSAR OS.
\section*{Acknowledgement}
\addcontentsline{toc}{section}{Acknowledgment}
This work was carried out on the StarBED computer cluster infrastructure
provided by the National Institute of Information and Communications
Technology (NICT).

\end{document}